\documentclass[twocolumn,pra,showpacs]{revtex4-1}
\usepackage{amssymb}
\usepackage{amsmath}
\usepackage{graphicx}
\usepackage{epsfig}

\begin{document}
\title{Su-Schrieffer-Heeger chain with one pair of $\mathcal{PT}$-symmetric defects}
\author{L. Jin}
\email{jinliang@nankai.edu.cn}
\author{P. Wang}
\author{Z. Song}
\affiliation{School of Physics, Nankai University, Tianjin 300071, China}
\begin{abstract}
The topologically nontrivial edge states induce $\mathcal{PT}$ transition in Su-Schrieffer-Heeger (SSH)
chain with one pair of gain and loss at boundaries. In this study, we investigated a pair of
$\mathcal{PT}$-symmetric defects located inside the SSH chain, in particular,
the defects locations are at the chain centre. The $\mathcal{PT}$ symmetry breaking of the bound states
leads to the $\mathcal{PT}$ transition, the $\mathcal{PT}$-symmetric phases and the localized states were studied.
In the broken $\mathcal{PT}$-symmetric phase, all energy levels break simultaneously in topologically trivial phase;
however, two edge states in topologically nontrivial phase are free from the influence of
the $\mathcal{PT}$-symmetric defects. We discovered $\mathcal{PT}$-symmetric bound states induced
by the $\mathcal{PT}$-symmetric local defects at the SSH chain centre. The $\mathcal{PT}$-symmetric bound states
significantly increase the $\mathcal{PT}$ transition threshold and coalesce to the topologically protected zero mode with vanishing probabilities on every other site of the left-half chain and the right-half chain, respectively.
\end{abstract}

\maketitle

\textbf{Introduction.}
The parity-time ($\mathcal{PT}$) symmetric non-Hermitian Hamiltonians can
possess real spectra~\cite%
{Bender,Dorey,Bender02,Ali,OL,Musslimani,Klaiman,GraefePRA} but nonunitary dynamics,
such as faster evolution~\cite{Faster,FasterObs} and power oscillation~\cite{CERuter}.
The $\mathcal{PT}$ system experiences a phase transition when its spectrum changes
between real and complex. The $\mathcal{PT}$ transition point is the exceptional point
associated with eigenstates coalescence~\cite{GraefePRA}. In one-dimensional system, the
exceptional point varies as system size and structure~\cite{Rotter15,SongQH}. The critical gain/loss rate approximately equals to the coupling strength
of a uniform chain for a pair of $\mathcal{PT}$-symmetric gain and loss
defects at boundary~\cite{LJin09}. For defects locations at center, the
critical gain/loss rate is the coupling strength between the two defects~%
\cite{YNJ}. Topological properties were extensively investigated in
condensed matter physics~\cite%
{Ryu,KM,Bernevig,Qi,Fu,Esaki,Kitaev,SChenPRL,GuoHM,WangHSciBulletin,WangH,ShiX}
and in photonic systems~\cite{Hafezi,IBloch,LSA}; furthermore, the $\mathcal{%
PT}$-symmetric topological insulator was proposed in two-dimensional (2D)
coupled optical resonators. Different with traditional Hermitian topological
insulator, the edge states are unidirectional amplified and damped~\cite%
{Chong}. Topological insulator states are $\mathcal{PT}$ symmetry breaking
in a $\mathcal{PT}$-symmetric non-Hermitian extension of Dirac Hamiltonians,
because the $\mathcal{PT}$ operator switches the edge states locations at
boundary~\cite{Hughes2011}. The Chern number was generalized for
non-Hermitian systems, the tight-binding model on the honeycomb and square
lattices under different symmetry classes were examined, broken $\mathcal{PT}
$-symmetric edge states with real part eigen energies being zero were found~%
\cite{Esaki2011}. The topologically chiral edge modes found in
2D non-Hermitian system were related to the exceptional point of the bulk Hamiltonian that characterized by two half-integer charges of the exceptional point~\cite{Leykam}.

The Su-Schrieffer-Heeger (SSH) chain~\cite{SSHPRL} with a pair of $\mathcal{%
PT}$-symmetric defects at boundary was studied, the edge states found in
topologically nontrivial phase are sensitive to the non-Hermiticity~\cite%
{SChen} and the critical non-Hermitian gain and loss approach zero~\cite{LS}%
. The non-Hermitian Kitaev and extended Kitaev
models were investigated similar as the SSH model~\cite{TongP,Klett}.
Optical systems are fruitful platforms for the investigation of $\mathcal{PT%
}$ symmetry~\cite{AGuo,Feng,PengNP,PengScience,Chang,Jing,JingSR}. The robust light interface states were discovered at the interface of a combined two SSH chains with different quantum phases of $\mathcal{PT}$ symmetry~\cite{FengLSR}. Recently,
non-Hermitian SSH chains were experimentally realized by coupled dielectric
microwave resonators~\cite{SchomerusOL,Schomerus} and photonic lattices~\cite%
{Zeuner,Szameit}. In passive SSH chains with periodical losses, single coupling disorder induces asymmetric topological zero mode~\cite{Schomerus} and $\mathcal{PT}$-symmetric topological zero mode interface states, respectively~\cite{Szameit}. $\mathcal{PT}$ symmetry breaking and topological properties were
theoretically investigated in other $\mathcal{PT}$-symmetric systems~\cite%
{Yuce14,Yuce15,YNJ2016}, the competition between two lattice defects can
induce $\mathcal{PT}$ symmetry breaking and restoration as non-Hermiticity
increasing in Aubry-Andr\'{e}-Harper model~\cite{YNJAA}.

In this work, we study an open SSH chain with one pair of $\mathcal{PT}$%
-symmetric gain and loss. The $\mathcal{PT}$-symmetric thresholds, the
topologically nontrivial edge states, and the local defects induced
$\mathcal{PT}$-symmetric bound states are investigated. The $\mathcal{PT}$-symmetry breaking
is closely related to the appearance of localized states. When considering the defects
located near the chain boundary, the edge states in topologically nontrivial
region break the $\mathcal{PT}$ symmetry if the defects are at the sites
with nonzero states distribution probabilities; otherwise, the edge states
are free from the influence of the on-site defects and the $\mathcal{PT}$
symmetry phase transition is induced by the bulk states: The
extended (bound) state induces the $\mathcal{PT}$ symmetry phase transition at weak (strong)
non-Hermiticity for defects near the chain boundary (at the chain center).

The $\mathcal{PT}$ transition threshold is the largest when the defects located at the chain center,
being the weak inhomogeneous couplings of the SSH chain. Two edge states
and four bound states exist at large non-Hermiticity, the number of breaking
energy levels increases as defects moving from the chain boundary to the
center. For defects near the chain center, when $\mathcal{PT}$ transition happens, all energy levels break simultaneously in
topologically trivial phase; by contrast, two topologically nontrivial edge
states are not $\mathcal{PT}$ symmetry breaking although in broken $\mathcal{%
PT}$ symmetry phase. The $\mathcal{PT}$ transition is associated with the
$\mathcal{PT}$ symmetry breaking of the eigenstate. The edge states and bound states probabilities
localize around the $\mathcal{PT}$-symmetric defects; therefore, they are $\mathcal{PT}$ symmetry breaking states except when the defects are the nearest neighbours. We discovered a pair of $\mathcal{PT}$-symmetric
bound states for the defects at the chain centre, the $\mathcal{PT}$%
-symmetric bound states significantly increase the $\mathcal{PT}$ transition
threshold, at which, the bound states coalesce to the topologically protected zero mode but their probabilities are not only confined to either the loss or the gain sublattice; the probabilities vanish on every other site of the left-half chain and the right-half chain, respectively.

\begin{figure}[tbp]
\centering
\includegraphics[bb=40 210 545 545, width=8.8 cm, clip]{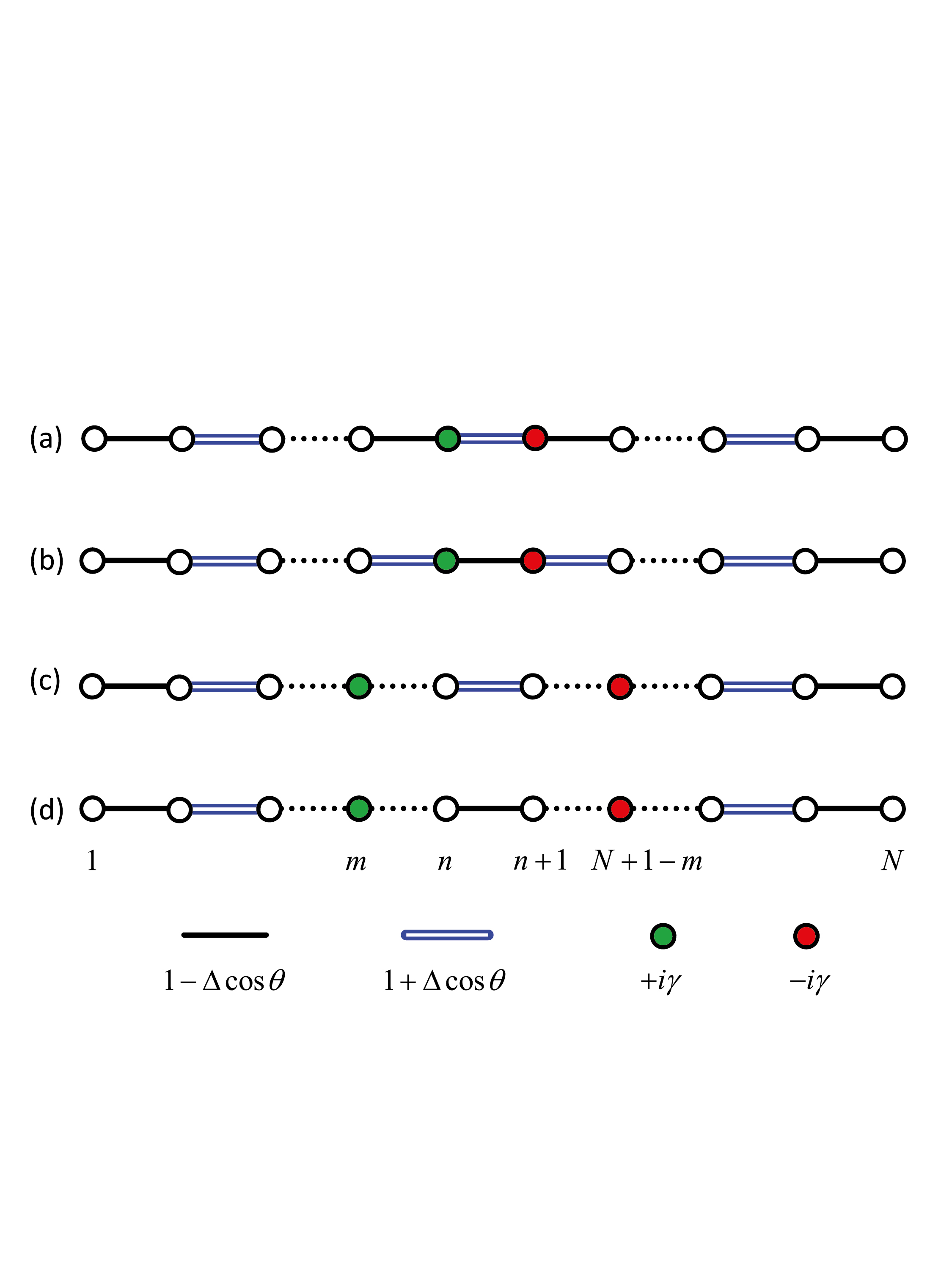}
\caption{\textbf{Schematic illustration of the $\mathcal{PT}$-symmetric SSH
chain.} The total chain number is even $N=2n$. The balanced gain and loss
are at site $m$ and its $\mathcal{P}$-symmetric site $N+1-m$. (\textbf{a})
Even $n$, $m=n$, (\textbf{b}) odd $n$, $m=n$, (\textbf{c}) even $n$, (%
\textbf{d}) odd $n$. The red and blue lines represent the inhomogeneous
couplings. The gain and loss sites are in green and red, respectively.}
\label{fig1}
\end{figure}

\textbf{$\mathcal{PT}$-symmetric non-Hermitian SSH chain.} In this section,
we introduce a one dimensional $N$-site SSH chain with one pair of $\mathcal{%
PT}$-symmetric imaginary defects, the system is schematically illustrated in
Fig.~\ref{fig1}. The couplings between neighbor sites are staggered $1\pm
\Delta \cos \theta $, which are modulated by parameter $\Delta $. The
coupled chain can be realized by optical resonators~\cite%
{ZhouZH,LeiFC,XiongH}. The defects pair includes a loss (in red) and a
balanced gain (in green)~\cite{OL,Musslimani,AGuo,CERuter,Jing,PengNP}. We
define $\mathcal{P}$ as the parity operator, which equals to a reflection
symmetry with respect to the chain center, satisfying $\mathcal{P}j\mathcal{P%
}^{-1}=N+1-j$. The time-reversal operator satisfies $\mathcal{T}i\mathcal{T}%
^{-1}=-i$. Under these definitions, the balanced gain and loss as on-site
defects pair satisfies the $\mathcal{PT}$ symmetry. The primary SSH
Hamiltonian $H_{0}$ is in form of $H_{0}=\sum_{j}^{N-1}[1+\left( -1\right)
^{j}\Delta \cos \theta ](c_{j}^{\dagger }c_{j+1}+c_{j+1}^{\dagger }c_{j})$,
where $c_{j}^{\dagger }$ ($c_{j}$) is the creation (annihilation) operator
on site $j$ for fermionic particles. The chiral symmetry protects the
topological properties of $H_{0}$ in topologically nontrivial region. In
this work, we confine our discussion within system with even $N$, the
non-Hermitian gain and loss defects are located at reflection symmetric
positions $m$ and $N+1-m$; the non-Hermitian extended SSH chain is%
\begin{equation}
H=H_{0}+i\gamma c_{m}^{\dagger }c_{m}-i\gamma c_{N+1-m}^{\dagger }c_{N+1-m}.
\label{H}
\end{equation}%
Note that $H$ satisfies the $\mathcal{PT}$ symmetry, and is expected to have
purely real spectrum. The analysis and conclusions are applicable to
corresponding bosonic particles~\cite{AAHPRL}.

Topologically nontrivial edge states disappear in system with universal
non-Hermiticity~\cite{Hughes2011,Esaki2011}, but remain in system with
localized non-Hermiticity~\cite{SChen,Yuce15}. \textit{The topology is
changed in the presence of universal non-Hermiticity, but is robust to
several impurities although the impurities are non-Hermitian. }The
traditional Hermitian SSH model has inhomogeneous staggered hoppings between
neighbor sites, the SSH Hamiltonian is a two-band model. Under periodical
boundary condition, the Berry phases of the two bands can be calculated,
both being $\pi $ in topologically nontrivial phase when $-\pi /2<\theta
<\pi /2$ and both being $0$ in topologically trivial phase when $-\pi
\leqslant \theta \leqslant -\pi /2$ and $\pi /2\leqslant \theta \leqslant
\pi $. Under open boundary condition, the bulk-edge correspondence indicates
the existence of edges states. For even $N$, two zero edge states appear in
region $-\pi /2<\theta <\pi /2$; by contrast, for odd $N$, single edge state
exists when $\theta \neq \pm \pi /2$~\cite{AAHPRL}. The edge states
probabilities are localized on the chain boundary, \textit{thus edge states
are immune to non-Hermitian on-site defects at the SSH chain center.}
\begin{figure*}[th]
\centering
\includegraphics[bb=0 0 535 280, width=18 cm, clip]{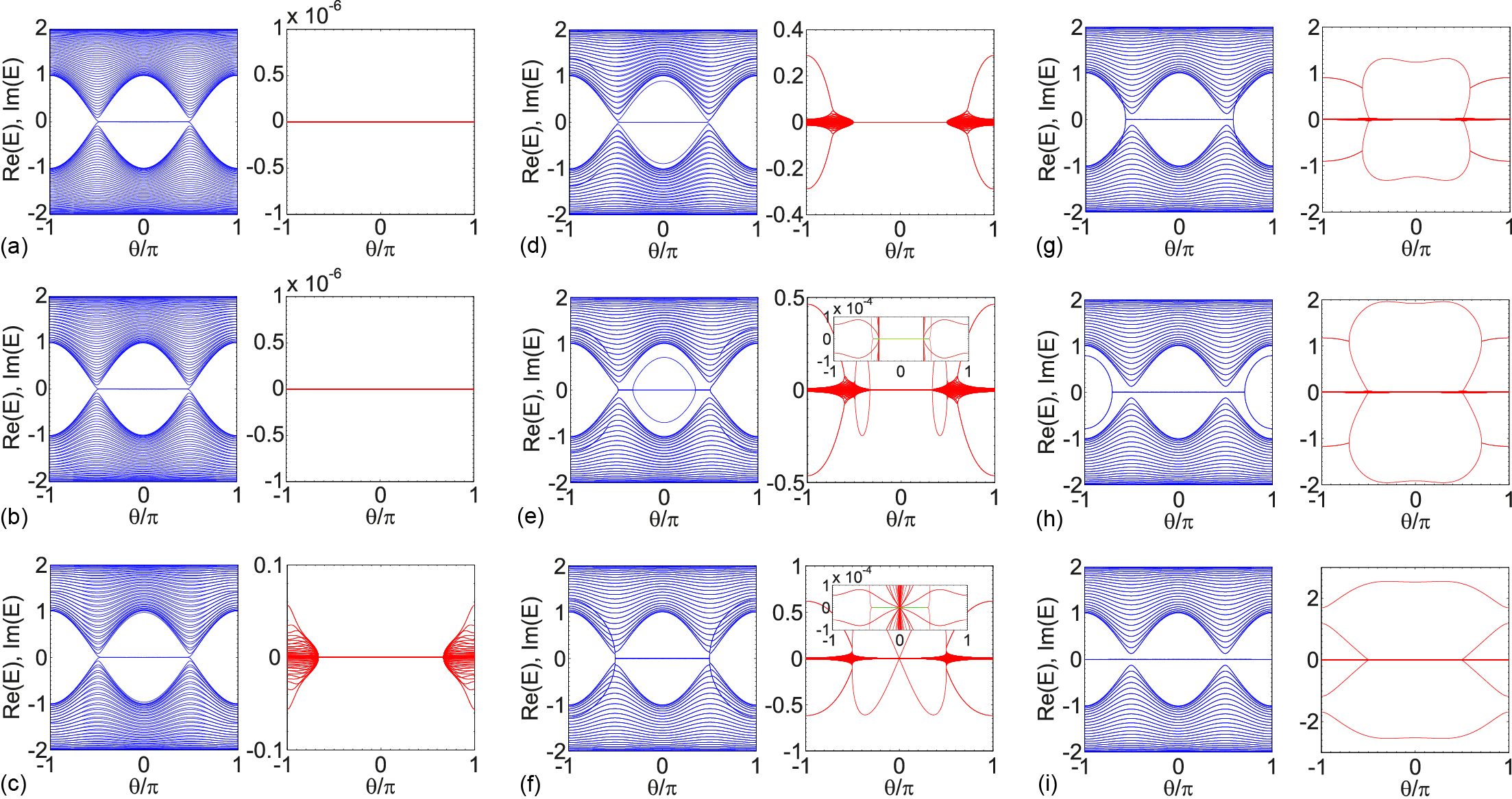}
\caption{\textbf{Spectrum under open boundary condition as a function of $%
\protect\theta $.} (\textbf{a}) $\protect\gamma =1/4$, (\textbf{b}) $\protect%
\gamma =1/2$, (\textbf{c}) $\protect\gamma =3/4$, (\textbf{d}), $\protect%
\gamma =1$, (\textbf{e}), $\protect\gamma =5/4$, (\textbf{f}) $\protect%
\gamma =3/2$, (\textbf{g}) $\protect\gamma =2$, (\textbf{h}) $\protect\gamma %
=5/2$, (\textbf{i}) $\protect\gamma =3$. Other system parameters are $N=100$%
, $\Delta =1/2$; the gain and loss are at sites $50$ and $51$.}
\label{fig2}
\end{figure*}

\textbf{$\mathcal{PT}$-symmetric phases.} We first consider $\mathcal{PT}$%
-symmetric SSH chain with even $n$ as follows (Fig.~\ref{fig1}a). The
defects are at the chain center ($m=n$). The SSH chain with odd $n$ has
similar results (Fig.~\ref{fig1}b). The Hamiltonian $H$ is always in the
exact $\mathcal{PT}$-symmetric phase as $\theta $ varies when $\gamma
<1-\Delta $; the exact $\mathcal{PT}$-symmetric region shrinks for $1-\Delta
<\gamma <1+\Delta $. At $\gamma >1+\Delta $, $H$ is in the broken $\mathcal{%
PT}$-symmetric phase for arbitrary $\theta $. In $\mathcal{PT}$ symmetry
broken phase, all energy levels are $\mathcal{PT}$ symmetry breaking in
topologically trivial region; however, two edge states are robust to gain
and loss defects in topologically nontrivial region, and can be composed
into a pair of $\mathcal{PT}$-symmetric states. As an illustration, we
numerically calculate the SSH chain spectra and depict them as a function of
$\theta $ for $N=100$ and $\Delta =1/2$ in Fig.~\ref{fig2}. The real and
imaginary parts of the spectra are plotted in blue and red lines,
respectively. Two topologically nontrivial edge states appear in the region $%
-\pi /2<\theta <\pi /2$.

Figures~\ref{fig2}a and~\ref{fig2}b show the entirely real spectrum as a
function of $\theta $ at $\gamma =1/4$ and $1/2$; however, the $\mathcal{PT}$%
-symmetric SSH chain is nondiagonalizable for $\gamma =1/2$ at $\theta =\pm
\pi $, where the coupling inhomogeneity is at maximum. The Hamiltonian $H$
become $n$ $2\times 2$ Jordan blocks after diagonalization, which indicates $%
n$ pairs of two states coalescence. For $\gamma >1/2$, the $\mathcal{PT}$
symmetry breaking appears at $\theta =\pm \pi $, the $\mathcal{PT}$ symmetry
of all eigen states breaks simultaneously. In Fig.~\ref{fig2}c, we depict
the SSH chain spectrum for $\gamma =3/4$. The exact $\mathcal{PT}$-symmetric
region is determined by $\gamma =1+\Delta \cos \theta $. Thus, the system is
in exact $\mathcal{PT}$-symmetric phase in region $-2\pi /3\leqslant \theta
\leqslant 2\pi /3$; in other regions, all the $N$ eigenvalues form $n$
conjugate pairs. The eigen states with the largest imaginary part (we refer
to the absolute values in the comparison) have highest probabilities
localized near the chain center. The inverse participation ratio (sum of
fourth power of the wave function amplitude $\sum_{j}\left\vert \psi
_{j}\right\vert ^{4}$) for the extended state scales as system size in order
of $N^{-1}$, but the IPR for the localized state approaches constant at
large system size. As $\gamma $ increasing, the probabilities are more
localized and form bound states at approximately $\gamma \gtrsim 1$. The
bound states are attributed to the $\mathcal{PT}$-symmetric non-Hermitian
impurities; and the bound states probabilities are localized near the chain
center.

In Fig.~\ref{fig2}d, $\gamma =1$, exact $\mathcal{PT}$-symmetric phase
shrinks to $-\pi /2\leqslant \theta \leqslant \pi /2$. In broken $\mathcal{PT%
}$-symmetric phase, the two red lines (two folders) with the maximum
imaginary parts in regions $-\pi \leqslant \theta \lesssim -0.7\pi $ and $%
0.7\pi \lesssim \theta \leqslant \pi $ correspond to four bound states.
Notably, the bound states eigenvalues change more rapidly than the extended
states and enter the extended states band around $\theta \approx \pm 0.7\pi $%
. As $\gamma $ increasing to $\gamma =5/4$ as shown in Fig.~\ref{fig2}e,
exact $\mathcal{PT}$-symmetric phase shrinks to $-\pi /3\leqslant \theta
\leqslant \pi /3$ (indicated by yellow line in the inset), where the SSH
chain has two degenerate zero edge states and two bound states with real
eigenvalues. The $\mathcal{PT}$ symmetry breaks out of region $-\pi
/3\leqslant \theta \leqslant \pi /3$. In the regions $-\pi /2\lesssim \theta
<-\pi /3$ and $\pi /3<\theta \lesssim \pi /2$, the SSH chain is in broken $%
\mathcal{PT}$-symmetric phase with two topologically nontrivial edge states.
There exist two bound states and two edge states, the two bound states have
pure imaginary eigenvalues with the largest imaginary part, exiting the
extended states band at $\theta \approx \pm \pi /2$ and become real for $%
\left\vert \theta \right\vert \leqslant \pi /3$. Notably, the two edge
states are also pure imaginary but with the smallest imaginary parts,
approaching zero as $\left\vert \theta \right\vert $ close to $0$. This is
because weak inhomogeneity for $\left\vert \theta \right\vert $ close to $%
\pi /2$ induces more spreading of edges states; by contrast, strong
inhomogeneity for $\left\vert \theta \right\vert $ close to $0$ induces more
localization of the edge states. In the numerical results, the imaginary
parts of edge states are negligible in the region $-0.4\pi \lesssim \theta
\lesssim 0.4\pi $ (indicated by green line in the inset), determined by the
coupling inhomogeneity $\Delta $. \textit{The two zero edge states are free
from the influence of non-Hermitian defects at (or close to) the chain
center.} In the regions $-\pi \leqslant \theta \leqslant -\pi /2$ and $\pi
/2\leqslant \theta \leqslant \pi $, there are four bound states with the
largest imaginary eigenvalues. In Fig.~\ref{fig2}f, the spectrum for $\gamma
=3/2$ is plotted. The bound states always have the largest imaginary
eigenvalues, the bound states are two folders out of region $-\pi /2<\theta
<\pi /2$, and one folder in region $-\pi /2<\theta <\pi /2$. For $\gamma
>3/2 $, the SSH chain spectrum is in the broken $\mathcal{PT}$-symmetric
phase at arbitrary $\theta $. \textit{The only real energy states are the
two topologically nontrivial edge states in the region} $-0.4\pi \lesssim
\theta \lesssim 0.4\pi $ (indicated by green line in the inset).

In Figs.~\ref{fig2}g and~\ref{fig2}f, $3/2<\gamma <2.96$, the imaginary
parts of the bound states experience a bifurcation in topological trivial
regions $-\pi \leqslant \theta \leqslant -\pi /2$ and $\pi /2\leqslant
\theta \leqslant \pi $. After bifurcation, the eigenvalues of the bound
states become pure imaginary, which is reflected from the real parts being
zero. In the real part of the energy spectrum, the zero is four folders, and
corresponds to the four bound states in the topologically trivial phase but
corresponds to two bound states and two edge states in the topologically
nontrivial phase. Notably, the four folders zero always exists in the real
part of energy spectrum for $\gamma >2.96$. The bound states have pure
imaginary eigenvalues. The bifurcation behavior disappears in the imaginary
parts of the bound states. In this situation, one pair of bound states have
larger imaginary eigenvalues than the other pair at arbitrary $\theta $.

In the situation of odd $n$ (Fig.~\ref{fig1}b), the coupling between the
gain and loss is $1-\Delta \cos \theta $; by contrast, this coupling is $%
1+\Delta \cos \theta $ for even $n$ case. The spectrum structures are
approximately the same as the even $n$ case shown in Fig.~\ref{fig2}, but
shifted by $\pi $ in parameter $\theta $, but two zero edge states still
exist in topologically nontrivial phase $-\pi /2<\theta <\pi /2$. In our
discussion of odd $n$ case, the coupling at the boundaries are unchanged as $%
1-\Delta\cos \theta$.

In the following, we discuss a general situation that the $\mathcal{PT}$%
-symmetric defects are inside the SSH chain ($0<m<n$) rather than at the SSH
chain center. The configuration is illustrated in Figs.~\ref{fig1}c and~\ref%
{fig1}d. The topological properties are robust to the one pair of gain and
loss defects; however, the $\mathcal{PT}$-symmetric properties change
significantly. The edge states with probabilities localized at the chain
boundary are free from the influence of non-Hermitian defects when they are
close to the chain center ($m\sim n$) at strong coupling inhomogeneity. By
contrast, for the gain and loss defects at the chain boundary ($m=1$), the $%
\mathcal{PT}$ symmetry of the SSH chain is fragile to the non-Hermitian
defects in the presence of topologically nontrivial edge states. This is
because the probabilities of edge states are the highest at the chain
boundary and decay exponentially. Thus, the influence of defects is the
greatest for edge states; any small gain and loss rate breaks the $\mathcal{%
PT}$ symmetry of the SSH chain in the topologically nontrivial region $-\pi
/2<\theta <\pi /2$; in topologically nontrivial region, the two edge states,
forming a conjugation pair with pure imaginary eigenvalues, are the only $%
\mathcal{PT}$ symmetry breaking states~\cite{SChen}.

As parameter $\theta $ varies, the number of breaking energy levels at
maximum appears in the topologically trivial phase. The number of breaking
energy levels at maximum is larger for larger $m$, which equals to $2m+2$
for odd $m\leqslant n-2$ in the topologically trivial phase but equals to $%
2m $ in the topologically nontrivial phase. By contrast, the number of
breaking energy levels at maximum is $2m$ for $m=n-1$ and $n$ in both
topologically trivial and nontrivial phases. When $m=n$, all energy levels
break simultaneously as shown in Fig.~\ref{fig2}c-e. \textit{The number of
breaking levels at maximum in topologically nontrivial region is two less
than that in topologically trivial region for }$m\leq n-2$\textit{\ at odd }$%
m$\textit{.}

For $N=100$, in topologically nontrivial phase of $\theta =0$, the coupling
inhomogeneity is the strongest. The number of $\mathcal{PT}$-symmetric
breaking energy levels at maximum is $2m$ for even $m<n-2$; which changes to
$2m-2$ ($2m$) for odd $25\leqslant m<n-2$ ($m<25$). The two edges states
have pure imaginary eigenvalue $|E_{\mathrm{ES}}|<10^{-10}$ for $m\geqslant
25$ along with defects moving from the chain boundary to the center, the
edge states can be considered as unaffected and the eigenvalues are zero for
odd $m\geqslant 25$. The critical gain/loss rate is $\gamma _{\mathrm{c}%
}\approx 0.07$ at $m=25$. At weak non-Hermiticity ($\gamma \ll 1$), bound
states disappear and two pairs (four) of extended states break first with
equal amount of energy imaginary parts. For $n-2\leqslant m\leqslant n$, the
number of $\mathcal{PT}$-symmetric breaking energy levels at maximum is $%
2m-2 $. The topologically nontrivial zero edge states are real valued.

\textbf{Edge states and bound states.} The defects support localized modes,
which can induce $\mathcal{PT}$ symmetry breaking~\cite{Bendix}. In the SSH
chain, the edge states break the $\mathcal{PT}$ symmetry if the gain and
loss are at the sites with nonzero distribution probabilities in
topologically nontrivial phase; otherwise, the edge states are free from the
influence of gain and loss and the $\mathcal{PT}$ symmetry phase transition
is induced by the bulk states, including the extended states induced $%
\mathcal{PT}$ symmetry phase transition at weak non-Hermiticity for defects
near the chain boundary and the bound states induced $\mathcal{PT} $
symmetry phase transition at strong non-Hermiticity for defects at the chain
center.

\begin{figure*}[t]
\centering
\includegraphics[bb=0 0 520 300, width=14 cm, clip]{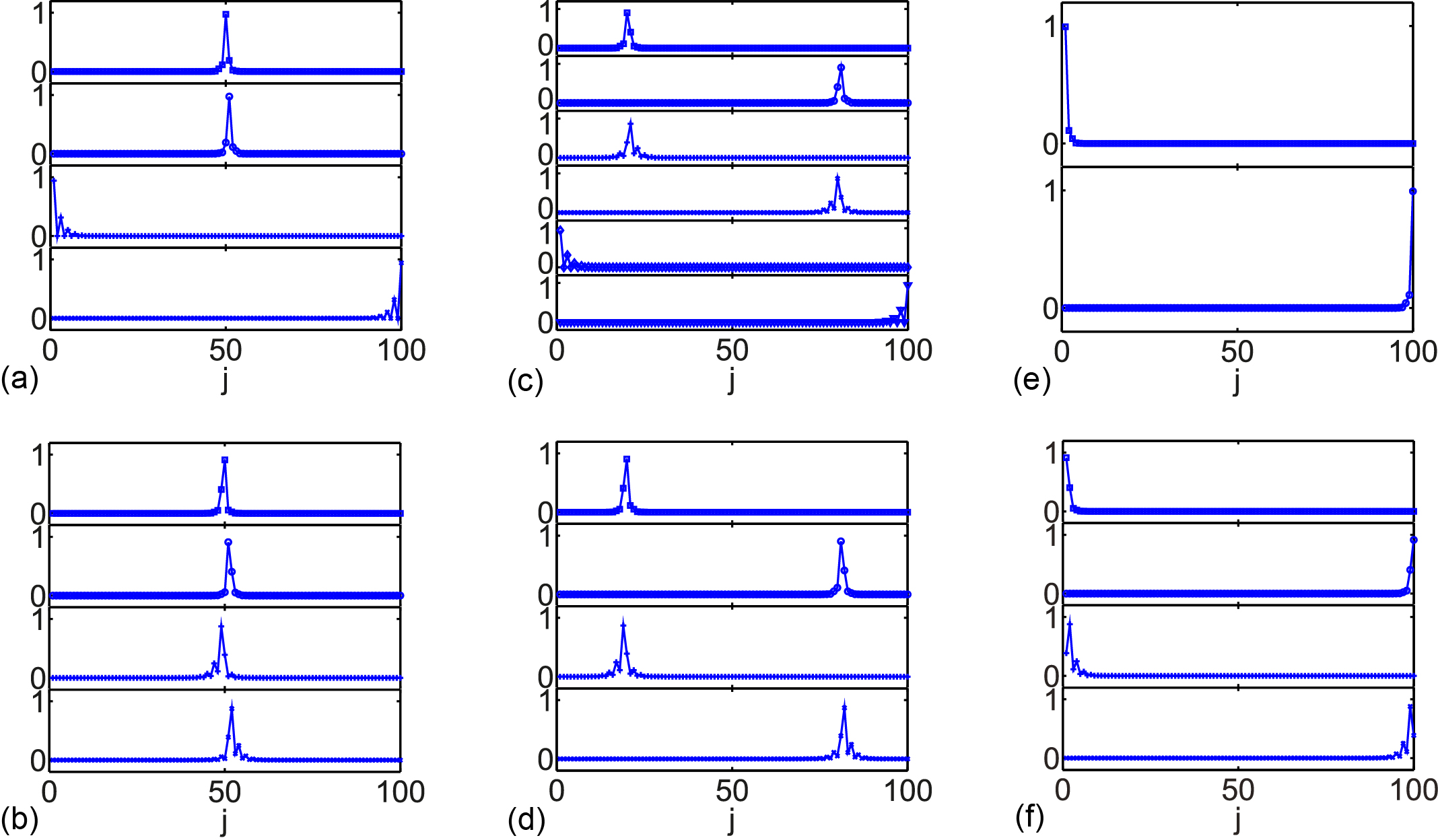}
\caption{\textbf{\ The localized edges states and bound states.} (\textbf{a,
b}) $m=50$, (\textbf{c, d}) $m=20$, (\textbf{e, f}) $m=1$. In the plots, $%
\protect\theta =0$ in (\textbf{a, c, e}), $\protect\theta =\protect\pi $ in (%
\textbf{b, d, f}). Other parameters are $\protect\gamma =4$, $N=100$, $%
\Delta =1/2$.}
\label{fig3}
\end{figure*}

The probabilities of two edge states for $\gamma =0$\ are staggered
decreasing from the chain boundary~\cite{AAHPRL}, the probability approaches
zero for every other site. The probabilities of two edge states on site $m$
and its $\mathcal{P}$-symmetric position $N+1-m$ are both zeros for even $m$%
. Thus, the edge states are unaffected (Figs.~\ref{fig3}a and~\ref{fig3}c).
By contrast, the influence of defects pair on the edge states is remarkable
for odd $m$, in particular, for the defects close to the chain boundary
(Fig.~\ref{fig3}e). The topologically nontrivial edge states are robust to
non-Hermitian defects, the nonvanishing distribution probabilities at the
chain boundary lead to $\mathcal{PT}$ symmetry breaking, and the energies of
two edge states become conjugate imaginary pair for small gain and loss~\cite%
{LS}.

At large non-Hermiticity, six localized states at maximum are found, which
include four bound states and two edge states as depicted in Fig.~\ref{fig3}%
c. The number of localized states depends on the non-Hermiticity and the
defects locations. For large enough gain and loss, all localized states
eigenvalues have zero real parts. In topologically trivial phase, the
localized states are four bound states; however, the number differs in
topologically nontrivial phase. Figure~\ref{fig3} depicts the edge states
and bound states probabilities for different defects locations. For $m=1$,
the defects at the chain boundary, two edge states are the only localized
states; for $2\leqslant m\leqslant n-1$, two edge states and four bound
states are found; for $m=n$, there are four localized states, including two
edge states and two bound states.

In Fig.~\ref{fig3}a, the localized states in topologically nontrivial phase
with gain and loss at the center are depicted. The localized states include
one conjugate pair of bound states and two edge states. The eigenvalues are $%
+3.6494i$, $-3.6494i$, and two degenerate $0$ from top to bottom. In Fig.~%
\ref{fig3}b, the SSH chain is in topologically trivial phase, the
eigenvalues of two pairs of bound states are $+3.2995i$, $-3.2995i$, $%
+0.6062i$, and $-0.6062i$ from the top to the bottom.

In Fig.~\ref{fig3}c, the eigenvalues of the four bound states are $+3.2594i$%
, $-3.2594i$, $+0.6136i$, and $-0.6136i$ from top to middle. The other two
states at bottom are the zero edge states. In Fig.~\ref{fig3}d, the
eigenvalues of four bound states are the same as those shown in Fig.~\ref%
{fig3}c, but with probabilities distributions slightly different. This is
because $\theta =0$ in Fig.~\ref{fig3}c but $\theta =\pi $ in Fig.~\ref{fig3}%
d, the two couplings $1\pm \Delta $ between each site and its nearest
neighbors switch in the two structures. The bound states probabilities are
localized near the defects and decay to zero at the chain boundary.
Therefore, the SSH chain structures at the boundary are not important for
the four bound states and they are approximately identical in the two cases
of $\theta =0$ and $\theta =\pi $. This conclusion is invalid when the gain
and loss defects locations are close, where bound states probabilities decay
from defects $i\gamma $ and $-i\gamma $ affected each other.

In case of defects at the chain boundary, the bound states disappear in
topologically nontrivial phase. In Fig.~\ref{fig3}e, the two edge states
eigenvalues are $+3.9445i$, $-3.9445i$, and being $\pm i\gamma $ for $%
n\rightarrow \infty $, both two edge states are fragile to impurities at the
chain boundary. The staggered decay of edge states disappears when $\gamma $
is large. In Fig.~\ref{fig3}f, the eigenvalues are $+3.3384i$, $-3.3384i$, $%
+0.5991i$, and $-0.5991i$. The staggered decay of state probabilities is
clearly seen for states with small imaginary eigenvalues (absolute values).

\begin{figure*}[t]
\centering
\includegraphics[bb=0 0 520 230, width=14 cm, clip]{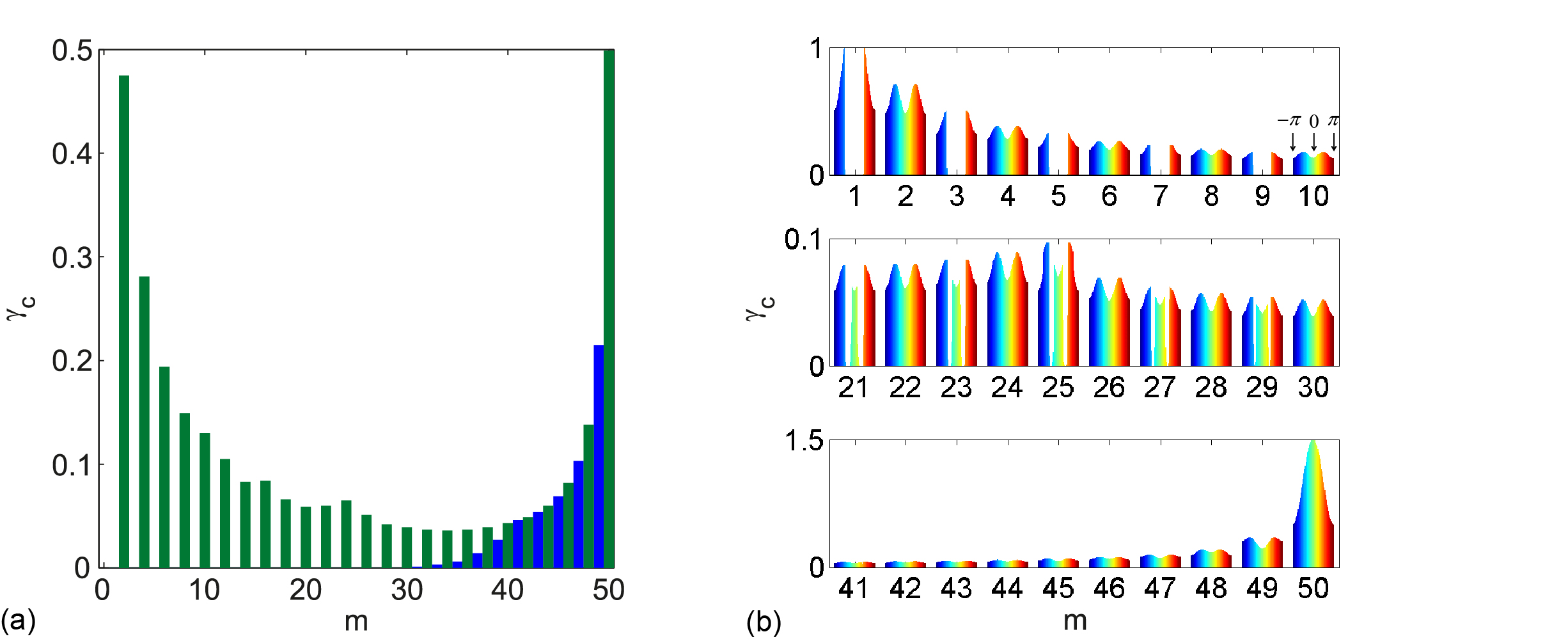}
\caption{\textbf{The numerically calculated $\protect\gamma _{\mathrm{c}}$
as a function of location $m$.} (\textbf{a}) $\protect\gamma _{\mathrm{c}}$
minimum in region $\protect\theta \in \left[ -\protect\pi ,\protect\pi %
\right] $ is depicted, which indicates the $\mathcal{PT}$-symmetric SSH
chain spectrum is entirely real for $\protect\gamma <\protect\gamma _{%
\mathrm{c}}$ at arbitrary $\protect\theta $. The blue bars are for odd $m$,
which approaches zero for $m<30$; the green bars are for even $m$. (\textbf{b%
}) $\protect\gamma _{\mathrm{c}}$ depicted for full $\protect\theta \in %
\left[ -\protect\pi ,\protect\pi \right] $, the dark blue from left to dark
red on the right represent $\protect\theta $ from $-\protect\pi $ to $%
\protect\pi $, the green area in the center corresponds to $\protect\theta %
=0 $, indicated by arrows in the upper right corner. Other parameters are $%
N=100 $, $\Delta =1/2$. }
\label{fig4}
\end{figure*}
The bound states with positive (negative) imaginary eigenvalues are centered
at the gain (loss) site. In the two pairs of bound states, the probabilities
decay faster for the one with larger imaginary eigenvalues. The
probabilities maxima of these two bound states are at the gain and loss
sites. The other pair of bound states has smaller imaginary eigenvalues,
which decay in a staggered way instead of monotonously and the decay is
slower. The probabilities maxima of this pair of bound states are at the
nearest neighbor site of the impurity, the site which has stronger coupling
strength between the impurity and its neighbors. As shown in Figs.~\ref{fig3}%
c and~\ref{fig3}d, the stronger couplings are $1+\Delta $ between $20$ ($81$%
) and $21$ ($80$) as shown in Fig.~\ref{fig3}c and $19$ ($82$) and $20$ ($81$%
) as\ shown in Fig.~\ref{fig3}d. The pair of bound states with smaller
positive (negative) imaginary parts are localized at $21$ ($80$) and $19$ ($%
82$), respectively. By contrast, for the gain and loss at the chain center,
bound states with smaller imaginary eigenvalues vanish (Fig.~\ref{fig3}a).
The situations are different for defects at the boundary ($m=1$) and the
center ($m=50$), the dimerized unit with stronger couplings is incomplete at
the boundary and the center in comparison with other cases, and the
localized states partially vanish accordingly.

In Fig.~\ref{fig4}a, we depict $\gamma _{\mathrm{c}}$ as a function of
location $m$. $\gamma _{\mathrm{c}}$ is maximal at $m=n$, being $1-\Delta $;
the minimum $\gamma _{\mathrm{c}}$ approaches zero for small odd $m$ (gain
and loss defects close to the chain boundary). The localized states are non-$%
\mathcal{PT}$-symmetric except for the two degenerate zero edge states,
which can be composed into $\mathcal{PT}$-symmetric form; and the real
valued bound states appear when defects are at the chain center. For even $m$%
, the edge states are unaffected; for odd $m$, the edge states break the $%
\mathcal{PT}$ symmetry when $m$ is small (defects near the chain boundary).
The $\mathcal{PT}$ symmetry is fragile to nonzero non-Hermiticity; when $m$
is large (defects near the chain center), the edge states are still
unaffected because the probabilities of edge states decayed to zero at the
locations of defects pair. In Fig.~\ref{fig4}a, $\gamma _{\mathrm{c}}$ is no
longer approaching $0$ for odd $m>30$ and monotonously increases as location
$m$ when $m>40$. These all reflect that the influence of defects pair on the
edge states is negligible and the two edge states energies are real and
still being zero. The bound states appear in conjugation pairs, being
non-degenerate; the bound states probabilities localize around each
impurity. $\mathcal{PT}$ symmetry is thus fragile to the bound states. An
exception is that when the defects are at the chain center ($m=50$), $%
\mathcal{PT}$-symmetric bound states can appear in topologically nontrivial
phase for $\gamma <\gamma _{\mathrm{c}}$ (in exact $\mathcal{PT}$-symmetric
phase). Figure~\ref{fig4}b depicts the contours of $\gamma _{\mathrm{c}}$ at
different location $m$ as function of $\theta $ in full region of $\theta
\in \left[ -\pi ,\pi \right] $. At $m=1$, $\gamma _{\mathrm{c}}$ maximum
equals to $1$ around $\left\vert \theta \right\vert =\pi /2$ and shapely
changes to zero in $\left\vert \theta \right\vert <\pi /2$, where
topologically nontrivial edge states appear. This is because that the edge
states are fragile to the on-site non-Hermitian gain and loss. Affected by
the edge states, the shape change of $\gamma _{\mathrm{c}}$ occurs near $%
\left\vert \theta \right\vert =\pi /2$ at odd $m$ for defects near the chain
boundary. The influence of edge states vanishes for defects near the chain
center and the bulk states induce $\mathcal{PT}$ symmetry phase transition.
Notably, $\gamma _{\mathrm{c}}$ increases dramatically at $m=50$ in
comparison with other cases. $\gamma _{\mathrm{c}}$ increases from $1-\Delta
=1/2$ at $\left\vert \theta \right\vert =\pi $ to $1+\Delta =3/2$ at $%
\left\vert \theta \right\vert =0$. This large $\mathcal{PT}$ transition
threshold implies the $\mathcal{PT}$-symmetric bound states may appear.

\textbf{The }$\mathcal{PT}$\textbf{-symmetric bound states. }In a large $N$
system, the bound states located in the chain center have amplitude decayed
to zero at the chain boundaries. At $m=n$, the bound states are analytically
calculated. The eigenenergy is $E=\left( t_{1}\chi +t_{2}\right) \cos \phi $%
, where $\sin \phi =-\gamma /t_{2}$, $\chi$ is the decay factor that depends on the chain configuration and the inhomogeneous couplings. The eigenenergy is real for $\left\vert
\gamma /t_{2}\right\vert \leqslant 1$. For even $n$, $t_{1}=1-\Delta \cos
\theta $, $t_{2}=1+\Delta \cos \theta $; for odd $n$, $t_{1}=1+\Delta \cos
\theta $, $t_{2}=1-\Delta \cos \theta $.

\begin{figure*}[t]
\centering
\includegraphics[bb=0 0 495 150, width=16 cm, clip]{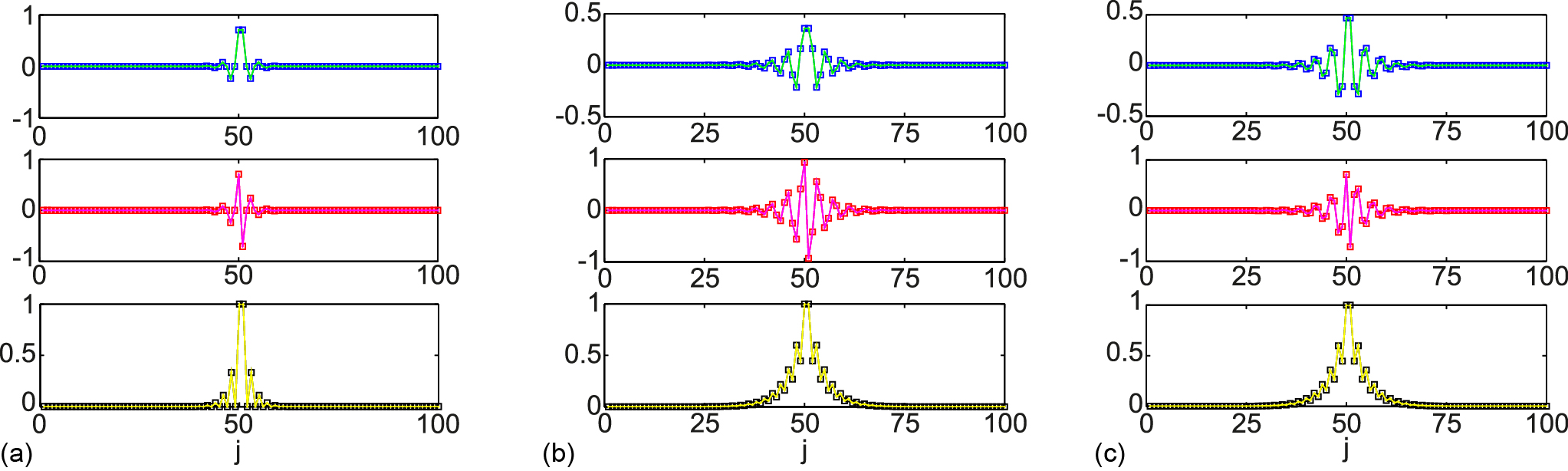}
\caption{\textbf{The $\mathcal{PT}$-symmetric bound states.} (\textbf{a})
The coalesced $\mathcal{PT}$-symmetric bound states with $E=0$ for $\protect%
\gamma=3/2$. One pair of $\mathcal{PT}$-symmetric bound states are shown for
$\protect\gamma=1$ in (\textbf{b, c}). (\textbf{b}) $E=2/\protect\sqrt{5}$, (%
\textbf{c}) $E=-2/\protect\sqrt{5}$. The upper, middle, and lower panels are
the real, imaginary and the absolute values of the wave function amplitude.
The square markers represent the numerical results for $N=100$, the solid
lines are the analytical results for $N\rightarrow \infty $. Other
parameters are $m=50$, $\protect\theta=0$, $\Delta=1/2$.}
\label{fig5}
\end{figure*}
Previously, the topologically protected $\mathcal{PT}$-symmetric zero modes were demonstrated in non-Hermitian SSH chains, the SSH chains globally possess balanced gain and loss in each dimerized unit cell; the topologically protected zero modes are interface states that induced by the coupling disorder in the SSH chain centre and the global loss~\cite{SchomerusOL,Schomerus,Szameit}. The interface states are confined to the passive sites with vanishing probability
distributions on the lossy sites~\cite{Schomerus}, and spread on both sublattices~\cite{Szameit}.
In our SSH chain with one pair of balanced gain and loss at the centre,
the existed $\mathcal{PT}$-symmetric bound states are different. \textit{The $\mathcal{PT}$-symmetric
bound states are induced by the non-Hermitian local defects and significantly enhance
the $\mathcal{PT}$ transition threshold.} The two bound states energies are
symmetric about zero energy and approach zero as the non-Hermiticity increases.
At the $\mathcal{PT}$ transition threshold, the bound states coalesce to the
$\mathcal{PT}$-symmetric zero mode, the zero mode is defective and topologically protected
by the band gap~\cite{Leykam}. The $\mathcal{PT}$-symmetric zero mode still differs with that found at the
interface between two topologically distinct $\mathcal{PT}$-symmetric lattices
induced by the coupling disorder~\cite{SchomerusOL,Schomerus}. The coalesced  $\mathcal{PT}$-symmetric zero mode probability vanishes on every other site of the left-half chain and the right-half chain, respectively. The two $\mathcal{PT}$-symmetric bound states are composed
by the edge state localized on the right edge $n$ of the left half chain and the edge
state localized on the left edge $n+1$ of the right half chain. When the coupling
strength between the neighbours at chain center (sites $n$ and $n+1$) is stronger
in the inhomogeneous couplings, the topologically protected zero mode appears
at $\gamma =\max (t_{1}$,$t_{2})$. The wave function contributions of the
on-site defects $\pm i\gamma $ and the coupling between the neighbour sites $%
n$ and $n+1$ cancel each other, they mimic a free-like boundary except for $%
\psi _{n}=i\psi _{n+1}$ ($\psi _{j}$ represents the wave function amplitude
for site $j$). The wave function amplitude stepped decays in form of $\chi
^{l}$, where $l$ is the dimerized unit cell index. The decay factor $\chi
=-t_{1}/t_{2}<1$ for even $n$ ($\chi =-t_{2}/t_{1}<1$ for odd $n$). In Fig.~%
\ref{fig5}, the $\mathcal{PT}$-symmetric bound states are depicted in
topologically nontrivial phase at $\theta =0$. The real parts (upper panels)
of the bound states are even functions of position while the imaginary parts
(middle panels) of the bound states are odd functions of position. In this
case ($t_{1}=1/2$, $t_{2}=3/2$), the $\mathcal{PT}$ transition threshold is
at $\gamma =3/2=t_{2}$, the $\mathcal{PT}$-symmetric bound states coalesce
and turn to the topologically protected zero mode; we depicted it in Fig.~%
\ref{fig5}a, the probability distribution vanishes for every other site of the left-half chain and the right-half chain, respectively. This topological zero mode differs with that found in the SSH chain with loss in each unit cell due to the distinct interface at the chain centre.~\cite{SchomerusOL,Schomerus}.

For $\left\vert \gamma /t_{2}\right\vert \neq 1$, the decay factor $\chi $
is
\begin{equation}
\chi =\frac{1}{2}\frac{t_{2}\sin ^{2}\phi }{t_{1}\cos ^{2}\phi }\left( 1-%
\sqrt{1+\left( \frac{t_{1}\cos ^{2}\phi }{t_{2}\sin ^{2}\phi }\right) ^{2}%
\frac{4}{\cos ^{2}\phi }}\right) .  \label{chi}
\end{equation}%
We choose the amplitude in the chain center $\left\vert \psi
_{n+1}\right\vert =1$ instead of renormalized the wave function for
convenience. In this situation, $\psi _{n+1}=e^{i\phi /2}$. The amplitude is
$[\psi _{n+1},\psi _{n+2},\psi _{n+3},\psi _{n+4},\cdots ]=e^{i\phi
/2}[1,\chi \cos \phi ,\chi ,\chi ^{2}\cos \phi ,\cdots ]$, the bound states are not confined to one sublattice similar as the previously found zero mode~\cite{Schomerus}; and the distributions of the $\mathcal{PT}$-symmetric bound states distinct at the centre due to the lattice difference~\cite{Szameit}. For $\gamma =1$
under $t_{1}=1/2$, $t_{2}=3/2$, we have $\cos \phi =\pm \sqrt{5}/3$; the
corresponding decay factors for the two $\mathcal{PT}$ symmetric bound
states are both $\chi =-0.6$ and the energies of the bound states are $E=\pm
2/\sqrt{5}$. The two $\mathcal{PT}$-symmetric bound states are depicted in
Fig.~\ref{fig5}b,c. Notably, they have identical probability distributions.

The wave function of the $\mathcal{PT}$-symmetric bound states at even $n$
is
\begin{widetext}
\begin{equation}
\psi _{j}=e^{i\sigma _{j}\phi /2}\{\frac{1+\chi \cos \phi }{2}+\left(
-1\right) ^{\left[ \sigma _{j}(j-n_{\mathrm{c}})\right] }\frac{1+\chi \cos
\phi }{2}\}\chi ^{\left[ \sigma _{j}(j-n_{\mathrm{c}})/2\right] },
\label{Psi}
\end{equation}%
\end{widetext}
where $n_{\mathrm{c}}=(N+1)/2$, the power exponent $\left[ \sigma _{j}(j-n_{%
\mathrm{c}})\right] $ of $(-1)$ in equation~\ref{Psi} represents the integer
part of $\sigma _{j}(j-n_{\mathrm{c}})$ and $\sigma _{j}$ is a sign function
defined as $\sigma _{j}=\mathrm{sgn}(j-n_{\mathrm{c}})$. At odd $n$ case,
the expression of $\psi _{j}$ is still valid; however, the bound states are
not real valued and $\mathcal{PT}$-symmetric bound states vanish. The values
$\phi $ for the bound states with complex eigenvalues are $\phi _{1}=\sin
^{-1}(-\gamma /t_{2})$, $\phi _{2}=-\sin ^{-1}(-\gamma /t_{2})$, $\phi
_{3}=\pi -\sin ^{-1}(-\gamma /t_{2})$, and $\phi _{4}=-\pi +\sin
^{-1}(-\gamma /t_{2})$; the corresponding decay factors $\chi $ and the wave
function of bound states can be obtained from equations~\ref{chi} and~\ref%
{Psi}.

\textbf{Conclusion.}
We have studied a pair of balance gain and
loss defects in a non-Hermitian SSH chain, the influence differs significantly as the
$\mathcal{PT}$-symmetric defects locations. The $\mathcal{PT}$ transition threshold has been
investigated, the number of broken energy levels at maximum increases as the
defects close to the chain center in the broken $\mathcal{PT}$-symmetric
phase; for the defects at the chain center, all energy levels break the $%
\mathcal{PT}$ symmetry simultaneously in topologically trivial phase, but
two edge states are free from $\mathcal{PT}$ symmetry breaking in
topologically nontrivial phase. When the defects are near the chain
boundaries, the edge states in topologically nontrivial phase break the $%
\mathcal{PT}$ symmetry if defects are at the sites with nonzero edge states
distribution probabilities; the $\mathcal{PT}$ symmetry breaking is caused
by the extended states at weak non-Hermiticity or by the bound states at
strong non-Hermiticity. The bound states probabilities are localized at the
defects and decay exponentially, thus are $\mathcal{PT}$-symmetric
breaking; however, the $\mathcal{PT}$-symmetric bound states can be formed
when the defects are at the SSH chain center, where the gain and loss are the
nearest neighbors. Therefore, the $\mathcal{PT}$ transition threshold in
this situation increases significantly, which is the largest and equals to
the weak inhomogeneous coupling. The $\mathcal{PT}$-symmetric bound states
are the topologically protected coalesced zero mode at the $\mathcal{PT}$ transition
threshold.

We acknowledge the support of National Natural Science Foundation of China
(Grant Nos. 11605094 and 11374163) and the Tianjin Natural Science
Foundation (Grant No. 16JCYBJC40800).

\end{document}